\documentclass[floatfix,twocolumn,showpacs,preprintnumbers,amsmath,amssymb,pra,superscriptaddress,longbibliography]{revtex4-1}
\usepackage{color}
\usepackage[usenames,dvipsnames,svgnames,table]{xcolor}
\usepackage[colorlinks=true,linkcolor=blue,urlcolor=blue,citecolor=blue]{hyperref}
\usepackage{mathtools}
\usepackage{graphicx}
\usepackage{dcolumn}
\usepackage{array}
\usepackage{lipsum}
\usepackage{bm}
\usepackage{subfigure}
\usepackage{amssymb}
\usepackage{multirow}
\usepackage{tabularx}
\usepackage{amsmath}
\usepackage{braket}
\usepackage{csquotes}
\graphicspath{{plots/}}
 \usepackage{lipsum}
\usepackage{mathrsfs}
\usepackage{MnSymbol}
	
\newcommand{\beq}{\begin{equation}}
\newcommand{\eeq}{\end{equation}}
\newcommand{\bea}{\begin{eqnarray}}
\newcommand{\eea}{\end{eqnarray}}

\begin{document}
\title{Re-weighting estimator for \textit{ab initio} path integral Monte Carlo simulations of fictitious identical particles}

\author{Tobias Dornheim}
\email{t.dornheim@hzdr.de}
\affiliation{Center for Advanced Systems Understanding (CASUS), D-02826 G\"orlitz, Germany}
\affiliation{Helmholtz-Zentrum Dresden-Rossendorf (HZDR), D-01328 Dresden, Germany}

\author{Pontus Svensson}
\affiliation{Center for Advanced Systems Understanding (CASUS), D-02826 G\"orlitz, Germany}
\affiliation{Helmholtz-Zentrum Dresden-Rossendorf (HZDR), D-01328 Dresden, Germany}

\author{Paul Hamann}
\affiliation{Center for Advanced Systems Understanding (CASUS), D-02826 G\"orlitz, Germany}
\affiliation{Helmholtz-Zentrum Dresden-Rossendorf (HZDR), D-01328 Dresden, Germany}
\affiliation{Institut f\"ur Physik, Universit\"at Rostock, D-18057 Rostock, Germany}

\author{Sebastian Schwalbe}
\affiliation{Center for Advanced Systems Understanding (CASUS), D-02826 G\"orlitz, Germany}
\affiliation{Helmholtz-Zentrum Dresden-Rossendorf (HZDR), D-01328 Dresden, Germany}

\author{Zhandos A.~Moldabekov}
\affiliation{Center for Advanced Systems Understanding (CASUS), D-02826 G\"orlitz, Germany}
\affiliation{Helmholtz-Zentrum Dresden-Rossendorf (HZDR), D-01328 Dresden, Germany}

\author{Panagiotis Tolias}
\affiliation{Space and Plasma Physics, Royal Institute of Technology (KTH), Stockholm, SE-100 44, Sweden}

\author{Jan Vorberger}
\affiliation{Helmholtz-Zentrum Dresden-Rossendorf (HZDR), D-01328 Dresden, Germany}

\begin{abstract}
The fermion sign problem constitutes one of the most fundamental obstacles in quantum many-body theory. Recently, it has been suggested to circumvent the sign problem by carrying out path integral simulations with a fictitious quantum statistics variable $\xi$, which allows for a smooth interpolation between the bosonic and fermionic limits  [\textit{J.~Chem.~Phys.}~\textbf{157}, 094112 (2022)]. This $\xi$-extrapolation method has subsequently been applied to a variety of systems and has facilitated the analysis of an x-ray scattering measurement taken at the National Ignition Facility with unprecedented accuracy [\textit{Nature Commun.}~\textbf{16}, 5103 (2025)]. Yet, it comes at the cost of performing an additional $10-20$ simulations, which, in combination with the required small error bars, can pose a serious practical limitation. Here, we remove this bottleneck by presenting a new re-weighting estimator, which allows the study of the full $\xi$-dependence from a single path integral Monte Carlo (PIMC) simulation. This is demonstrated for various observables of the uniform electron gas and also warm dense beryllium. We expect our work to be useful for future PIMC simulations of Fermi systems, including ultracold atoms, electrons in quantum dots, and warm dense quantum plasmas.
\end{abstract}
\maketitle

\section{Introduction\label{sec:introduction}}

The \emph{ab initio} path integral Monte Carlo (PIMC) method~\cite{Fosdick_PR_1966,Takahashi_Imada_PIMC_1984,Berne_JCP_1982} constitutes one of the most successful theoretical tools for the study of finite temperature quantum many-body systems in thermal equilibrium. Indeed, it is possible to carry out exact simulations of up to $N\sim10^3-10^4$ interacting bosons~\cite{boninsegni1,boninsegni2} (and also hypothetical distinguishable quantum particles, often denoted as \emph{boltzmannons} in the literature~\cite{Clark_PRL_2009,Dornheim_CPP_2016}) without the need for any empirical external input such as the exchange--correlation functional in density functional theory.

In stark contrast, \emph{ab initio} PIMC simulations of interacting fermions are afflicted with the notorious fermion sign problem~\cite{troyer,Loh_PRB_1990,dornheim_sign_problem}, which directly follows from the anti-symmetry of the corresponding thermal density matrix under the exchange of particle coordinates. More specifically, the sign problem manifests as an exponential bottleneck with respect to the system size $N$ or the inverse temperature $\beta$, thus severely limiting the application of PIMC for fermionic systems. This is highly unsatisfactory, as fermions are known to exhibit a wealth of interesting phenomena, including the BCS transition of ultracold atoms such as $^3$He at low temperatures~\cite{Kagan_2019}, the formation of roton-type excitations in various quantum liquids~\cite{Godfrin2012,dornheim_dynamic,Dornheim_Nature_2022,koskelo2023shortrange,Takada_PRB_2016,Chuna_JCP_2025}, as well as the formation of Wigner molecules~\cite{PhysRevLett.82.3320} and Wigner crystals~\cite{Filinov_PRL_2001} of electrons in low-density quantum dots. A particularly active field of research is given by the study of warm dense quantum plasmas~\cite{vorberger2025roadmapwarmdensematter,wdm_book}. These are characterized by the complex interplay of Coulomb coupling, quantum degeneracy and delocalization, strong thermal excitations and often also partial ionization~\cite{new_POP,wdm_book,vorberger2025roadmapwarmdensematter,Dornheim_review}, making their rigorous theoretical description particularly challenging. In nature, such warm dense matter (WDM) conditions (see Sec.~\ref{sec:parameters} below) abound in a variety of astrophysical systems such as giant planet interiors~\cite{Benuzzi_Mounaix_2014,Guillot2018}, brown dwarfs~\cite{becker} and white dwarf atmospheres~\cite{SAUMON20221}. In addition, both the fusion fuel and the surrounding ablator material have to traverse the WDM regime in a controlled way to reach ignition~\cite{hu_ICF} in laser fusion experiments~\cite{Betti2016,Hurricane_RevModPhys_2023,drake2018high}. Despite recent spectacular achievements~\cite{Zylstra2022,AbuShawareb_PRL_2024,Williams2024}, the further optimization of experimental set-ups that is needed to develop inertial fusion energy into a reliable future source of clean energy~\cite{Batani_roadmap} will require the development of integrated multi-scale simulations with real predictive capability, for which, in turn, a more rigorous theoretical understanding of WDM microphysics is indispensable~\cite{vorberger2025roadmapwarmdensematter}.

Due to this pressing need to understand interacting quantum Fermi systems, there has been a remarkable surge of developments in the field of corresponding quantum Monte Carlo simulations at finite temperatures~\cite{Brown_PRL_2013,Schoof_PRL_2015,Blunt_PRB_2014,Chin_PRE_2015,Malone_JCP_2015,Malone_PRL_2016,Dornheim_NJP_2015,Groth_PRB_2016,Joonho_JCP_2021,Hirshberg_JCP_2020,Dornheim_JCP_2020,Xiong_JCP_2022,Xiong_PRE_2024,Dornheim_JCP_xi_2023,Dornheim_JPCL_2024,Dornheim_NatComm_2025,Hou_PRB_2022,Militzer_PRE_2021,Driver_PRL_2012,Driver_PRE_2018,Yilmaz_JCP_2020,Hunger_PRE_2021,Dornheim_PRB_ChemPot_2025}, see also the recent review articles by Dornheim \emph{et al.}~\cite{Dornheim_review} and Bonitz \emph{et al.}~\cite{Bonitz_POP_2024}, as well as the WDM roadmap by Vorberger \emph{et al.}~\cite{vorberger2025roadmapwarmdensematter}. A particularly promising idea has been suggested by Xiong and Xiong~\cite{Xiong_JCP_2022} for path integral molecular dynamics (PIMD), who have proposed to introduce a fictitious quantum statistics variable $\xi\in[-1,1]$ that is continuous, where the cases of $\xi=-1,0,1$ correspond to the Fermi-Dirac, Maxwell-Boltzmann, and Bose-Einstein statistics, respectively. The basic idea is to extrapolate expectation values from the sign-problem free domain of $\xi\geq0$ to the fermionic limit of $\xi=-1$ using an empirical polynomial ansatz. This method has subsequently been adapted from PIMD to PIMC in Ref.~\cite{Dornheim_JCP_xi_2023}, and was successfully applied to a variety of physical systems, including electrons in quantum dots~\cite{Xiong_JCP_2022,Dornheim_JCP_xi_2023}, ultracold $^3$He~\cite{morresi2024normalliquid3hestudied} and a variety of WDM systems~\cite{Dornheim_JPCL_2024,Dornheim_JCP_2024,Dornheim_MRE_2024,Dornheim_POP_2025,schwalbe2025staticlineardensityresponse,Dornheim_NatComm_2025,morresi2025studyuniformelectrongas,dornheim2025fermionicfreeenergiestextitab,svensson2025acceleratedfreeenergyestimation}. The great strength of this $\xi$-extrapolation scheme is that it basically removes the exponential computational bottleneck of the fermion sign problem with respect to the system size for weak to moderate levels of quantum degeneracy, facilitating simulations of up to $N=1000$ electrons~\cite{Dornheim_JPCL_2024,svensson2025acceleratedfreeenergyestimation} and the application of PIMC for the interpretation of x-ray scattering experiments as it has recently been demonstrated for strongly compressed beryllium realized at the National Ignition Facility (NIF) in Livermore, USA~\cite{Dornheim_NatComm_2025,Dornheim_POP_2025,schwalbe2025staticlineardensityresponse}.
Yet, the favorable scaling of the $\xi$-extrapolation method with respect to the sign problem comes at the cost of having to perform PIMC simulations at $N_\xi=10-20$ different $\xi$ values, which becomes a serious bottleneck at comparably large system sizes.

Here, we remove this bottleneck by introducing a new re-weighting estimator, which allows evaluation of the full $\xi$-dependence from a single PIMC simulation of an, in principle, arbitrary reference value $\xi_\textnormal{ref}$. To demonstrate the versatility of this idea, we consider two representative examples. First, we simulate the uniform electron gas (UEG)~\cite{review,quantum_theory,loos} at a high density and find perfect agreement with previous PIMC calculations with the standard $\xi$-extrapolation technique~\cite{Dornheim_JCP_xi_2023} as well as with highly accurate configuration PIMC reference data~\cite{Dornheim_PRB_2016,Groth_PRB_2016}. Second, we present new results for strongly compressed beryllium as it has recently been realized at the NIF~\cite{Tilo_Nature_2023,Dornheim_NatComm_2025}, thus highlighting the considerable potential of the present set-up for important real world applications. Our method is very general and can be applied to a broad range of Fermi systems including ultracold atoms~\cite{morresi2024normalliquid3hestudied,Dornheim_SciRep_2022,Dornheim_PRA_2020,Ceperley_PRL_1992} and electrons in quantum dots~\cite{Reimann_RevModPhys_2002,Xiong_JCP_2022,Dornheim_NJP_2022}.

The paper is organized as follows: in Sec.~\ref{sec:theory}, we provide the required theoretical background, including a brief discussion of dimensionless parameters (\ref{sec:parameters}), the basic idea of the PIMC method (\ref{sec:PIMC}), a concise overview of the simulation of fictitious identical particles and the new re-weighting estimator (\ref{sec:fictitious}).
In Sec.~\ref{sec:results}, we show new simulation results for the warm dense electron gas (\ref{sec:UEG}) and for strongly compressed beryllium (\ref{sec:beryllium}). The paper is concluded by a summary and outlook in Sec.~\ref{sec:outlook}.

\section{Theory\label{sec:theory}}

We assume Hartree atomic units throughout this work.

\subsection{Dimensionless parameters\label{sec:parameters}}

It is convenient to characterize WDM systems by (at least) two dimensionless parameters~\cite{Ott2018}. The density parameter $r_s=(3/4\pi n_e)^{1/3}$, where $n_e=N_e/\Omega$ is the electronic number density ($N_e$ the total number of electrons,  $\Omega=L^3$ the volume of the cubic simulation cell) also serves as the quantum coupling parameter. For a constant level of quantum degeneracy, the uniform electron gas attains the limit of a non-interacting ideal Fermi gas in the high density limit of $r_s\to0$, whereas it forms a strongly coupled electron liquid~\cite{dornheim_electron_liquid,dornheim_dynamic,Chuna_JCP_2025,Tolias_JCP_2021,Tolias_JCP_2023} and eventually a Wigner crystal~\cite{Azadi_Wigner_2022,Drummond_PRB_Wigner_2004} at large $r_s$. The electron liquid regime in particular has attracted considerable recent interest due to the emergence of a roton-type feature in the dynamic structure factor $S_{ee}(\mathbf{q},\omega)$ at intermediate wavenumbers $q\sim2q_\textnormal{F}$~\cite{dornheim_dynamic,Dornheim_Nature_2022,chuna2025estimatesdynamicstructurefactor,Takada_PRB_2016,Takada_PRL_2002,koskelo2023shortrange}, with $q_\textnormal{F}=(9\pi/4)^{1/3}/r_s$ being the Fermi wavenumber~\cite{quantum_theory}; it has been explained consistently in terms of the alignment of pairs of electrons~\cite{Dornheim_Nature_2022,Dornheim_Force_2022} and by the formation of excitons~\cite{Takada_PRB_2016,koskelo2023shortrange}, and is also predicted to be measurable in warm dense hydrogen~\cite{Hamann_PRR_2023}. The second important parameter is given by the degeneracy temperature $\Theta=k_\textnormal{B}T/E_\textnormal{F}$, with $E_\textnormal{F}=q_\textnormal{F}^2/2$ the Fermi energy. Specifically, $\Theta\gg1$ indicates the semi-classical limit~\cite{Dornheim_HEDP_2022,Roepke_PRE_2024} where quantum delocalization and, in particular, quantum degeneracy effects play a minor role, whereas $\Theta\ll1$ indicates the quantum degenerate ground-state limit.

In the WDM regime, we have $\Theta\sim r_s\sim1$, leading to the well-known interplay of Coulomb coupling, quantum delocalization and degeneracy, thermal excitations, and often also partial ionization. The absence of a small parameter necessitates a holistic treatment of all of these effects, which is challenging even for state-of-the-art \emph{ab initio} methods~\cite{new_POP,wdm_book,vorberger2025roadmapwarmdensematter}. The key advantage of PIMC is that it is, in principle, capable of providing exact results without the need for empirical external input, thus opening up the way for reliable benchmark data sets and accurate easy-to-use parametrizations thereof~\cite{ksdt,groth_prl,dornheim_ML,Dornheim_PRB_ESA_2021,Dornheim_PRL_2020_ESA}.

\subsection{Path integral Monte Carlo\label{sec:PIMC}}

Due to the extensive body of literature on the PIMC method~\cite{cep,Berne_JCP_1982,Pollock_PRB_1984,boninsegni1}, we here restrict ourselves to a concise discussion of the key relations that are of relevance to the present work.

A fundamental quantity in statistical physics is given by the partition function, which we express in the compact form
\begin{eqnarray}\label{eq:Z_generic}
    Z_\xi(\beta,N,\Omega) = \sum_{\sigma\in S_N} \int\textnormal{d}\mathbf{X}\ W_\xi(\mathbf{X})\ ,
\end{eqnarray}
with the definition
\begin{eqnarray}
    W_\xi(\mathbf{X}) = W(\mathbf{X})\ \xi^{N_\textnormal{pp}}\ .
\end{eqnarray}
Specifically, we consider a canonical ensemble where the inverse temperature $\beta=1/k_\textnormal{B}T$, particle number $N$ and volume $\Omega=L^3$ are fixed. In the imaginary-time path integral representation of statistical mechanics, each particle is mapped onto a path through the imaginary time $t=-i\tau$ with $\tau\in[0,\beta]$ (throughout the remainder of this work, we will follow the usual convention of simply denoting $\tau$ as "imaginary time", although technically it is neither) which is discretized into $P$ time slices of length $\epsilon=\beta/P$. The basic idea of the PIMC method is to use an efficient adaption of the acclaimed Metropolis algorithm~\cite{metropolis} to sample all possible such path configurations $\mathbf{X}$ which have to be taken into account according to their configuration weight $W(\mathbf{X})$, which is strictly positive and contains both contributions from the kinetic and potential energy operators $\hat{K}$ and $\hat{V}$.
Finally, the variable $\xi$ controls the quantum statistics (where $N_\textnormal{pp}$ is the number of pair exchanges of a given configuration), with $\xi=-1,0,1$ corresponding to fermions, boltzmannons and bosons, respectively, and which is taken into account by the explicit summation over all possible permutations $\sigma$ of the corresponding permutation group $S_N$~\cite{Dornheim_permutation_cycles}. Note that we have assumed a single particle species in Eq.~(\ref{eq:Z_generic}) for simplicity, with the generalization to multiple species being straightforward.

Let us next consider the expectation value of an arbitrary observable $\hat{O}$,
\begin{eqnarray}\label{eq:expectation_value}
    \braket{\hat O}_\xi = \frac{1}{Z_\xi}\sum_{\sigma\in S_N}\int\textnormal{d}\mathbf{X}\ W_\xi(\mathbf{X}) O(\mathbf{X)}\ ,
\end{eqnarray}
where $O(\mathbf{X})$ denotes the corresponding Monte Carlo estimator. Note that $\beta$, $N$, and $\Omega$ are being suppressed throughout the remainder of this work for simplicity.
For $\xi<0$, $W_\xi(\mathbf{X})$ can be both positive and negative depending on $N_\textnormal{pp}$. Since Monte Carlo sampling probabilities must not be negative, one usually samples configurations $\mathbf{X}$ according to the absolute value of $W_\xi(\mathbf{X})$, which is simply given by $W_{|\xi|}(\mathbf{X})$. The exact sign-afflicted expectation value is then given by
\begin{eqnarray}\label{eq:ratio}
    \braket{\hat O}_{\xi<0} = \frac{\braket{\hat{O}\hat{S}}_{|\xi|}}{\braket{\hat{S}}_{|\xi|}} \ .
\end{eqnarray}
Here $S(\mathbf{X})=W_\xi(\mathbf{X})/W_{|\xi|}(\mathbf{X})$ corresponds to the sign of the configuration weight of a configuration $\mathbf{X}$ and the denominator of Eq.~(\ref{eq:ratio}) is usually denoted as the \emph{average sign}~\cite{troyer,Loh_PRB_1990,dornheim_sign_problem}.  It constitutes a convenient measure for the degree of cancellation of positive and negative terms in the PIMC simulation, with simulations generally being feasible for $S = \braket{\hat{S}}_\xi\gtrsim10^{-2}-10^{-3}$. Importantly, the average sign generally decreases exponentially with increasing system size or decreasing temperature. The resulting exponential computational bottleneck is known as the \emph{fermion sign problem}, which is one of the most fundamental obstacles in theoretical physics, quantum chemistry, material science, and related disciplines.

\subsection{Fictitious identical particles and re-weighting\label{sec:fictitious}}

Recently, Xiong and Xiong~\cite{Xiong_JCP_2022} have suggested to circumnavigate the sign problem by carrying out path integral simulations for non-integer values of $\xi\in[-1,1]$, corresponding to fictitious identical particles. In particular, they have argued that the expectation value of any given observable $O(\xi)=\braket{\hat O}_\xi$ is an analytical function with respect to $\xi$, which might allow one to extrapolate from the sign-problem free domain of $\xi\geq0$ to the fermionic limit of $\xi=-1$. In practice, this has been achieved using the empirical parabolic ansatz
\begin{eqnarray}\label{eq:extrapolation}
    O(\xi) = a_O + b_O\xi + c_O\xi^2\ ,
\end{eqnarray}
where $a_O$, $b_O$, and $c_O$ are the free fit parameters. Generally, it has been found that using $N_\xi=10-20$ $\xi$-values allows for a reliable extrapolation for weak to moderate degrees of quantum degeneracy, see Refs.~\cite{Dornheim_JCP_xi_2023,Dornheim_JPCL_2024,Dornheim_NatComm_2025,Dornheim_JCP_2024}.
On the one hand, this $\xi$-extrapolation technique can be based exclusively on PIMC simulations in the sign-problem free domain of $\xi\geq0$, which are not subject to the exponential bottleneck with respect to the system size inherent to the sign problem~\cite{Dornheim_JPCL_2024}.
On the other hand, the requirement for $\sim20$ independent PIMC simulations at a large system size can become cost prohibitive by itself, thus sparking the need for further methodological developments.


It is straightforward to see that one can re-cast the expectation value of any given $\xi$ in terms of expectation values at another reference value $\xi_\textnormal{ref}$,
\begin{widetext}
\begin{eqnarray}
\braket{\hat{O}}_\xi = \frac{Z_{\xi_\textnormal{ref}}}{Z_\xi}  \frac{1}{Z_{\xi_\textnormal{ref}}} \sum_{\sigma\in S_N}\int\textnormal{d}\mathbf{X}\ W_{\xi_\textnormal{ref}}(\mathbf{X}) \underbrace{ \frac{W_\xi(\mathbf{X})}{W_{\xi_\textnormal{ref}}(\mathbf{X})}  O(\mathbf{X}) }_{O_{\xi_\textnormal{ref},\xi}(\mathbf{X})} = \frac{Z_{\xi_\textnormal{ref}}}{Z_\xi} \braket{\hat{O}_{\xi_\textnormal{ref},\xi}}_{\xi_\textnormal{ref}} \ ,
\end{eqnarray}
\end{widetext}
where the ratio of partition functions is 
\begin{widetext}
\begin{eqnarray}
\frac{Z_\xi}{Z_{\xi_\textnormal{ref}}} = \frac{1}{Z_{\xi_\textnormal{ref}}} \sum_{\sigma\in S_N}\int\textnormal{d}\mathbf{X}\ W_{\xi_\textnormal{ref}}(\mathbf{X}) \frac{W_\xi(\mathbf{X})}{W_{\xi_\textnormal{ref}}(\mathbf{X})} = \left< \frac{W_\xi(\mathbf{X})}{W_{\xi_\textnormal{ref}}(\mathbf{X})} \right>_{\xi_\textnormal{ref}} \ ,
\end{eqnarray}
\end{widetext}
leading to the final estimator
\begin{eqnarray}
\Rightarrow \braket{\hat O}_\xi = \frac{ \braket{\hat{O}_{\xi_\textnormal{ref},\xi}}_{\xi_\textnormal{ref}} }{ \left< \frac{W_\xi(\mathbf{X})}{W_{\xi_\textnormal{ref}}(\mathbf{X})} \right>_{\xi_\textnormal{ref}} } \ . \label{eq:final}
\end{eqnarray}
Eq.~(\ref{eq:final}) constitutes the central result of the present work with the ratio of the configuration weights simply given by
\begin{eqnarray}\label{eq:re}
    \frac{W_\xi(\mathbf{X})}{W_{\xi_\textnormal{ref}}(\mathbf{X})} = \left(\frac{\xi}{\xi_\textnormal{ref}}\right)^{N_\textnormal{pp}}\ .
\end{eqnarray}
The estimator of the observable $\hat{O}$ is then re-weighted by Eq.~(\ref{eq:re}) and subsequently normalized by the corresponding ratio of partition functions.

\section{Results\label{sec:results}}

All results presented in this work have been obtained using a canonical extended ensemble adaption~\cite{Dornheim_PRB_nk_2021} of the worm algorithm by Boninsegni \emph{et al.}~\cite{boninsegni1,boninsegni2} that has been implemented into the open-source \texttt{ISHTAR} code~\cite{ISHTAR}. Our new PIMC simulation data are available online in a freely accessible repository~\cite{repo}. Note that we consider spin-unpolarized systems with $N_\uparrow=N_\downarrow=N/2$ throughout.

\subsection{Uniform electron gas\label{sec:UEG}}

\begin{figure}
    \centering
    \includegraphics[width=1\linewidth]{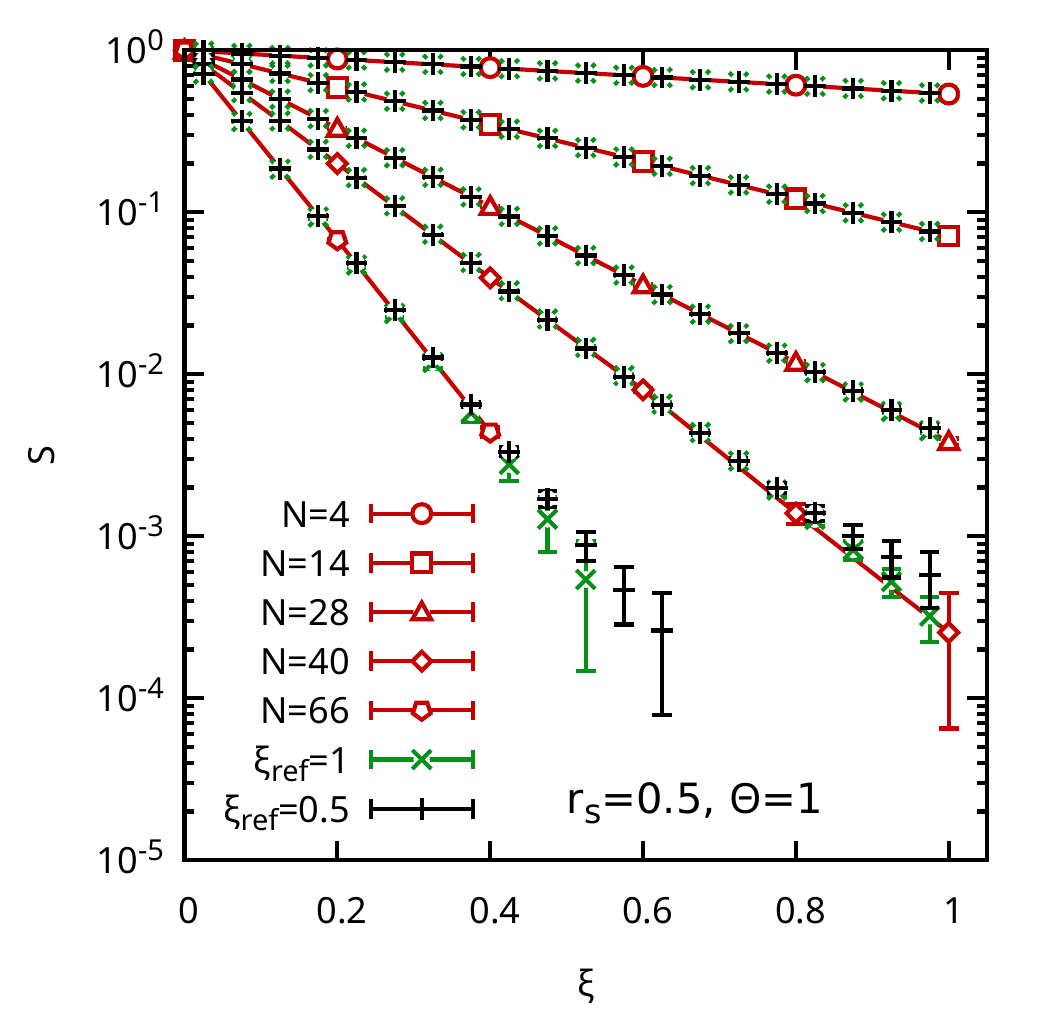}
    \caption{\label{fig:UEG_sign} Average sign $S$ of the unpolarized UEG at $r_s=0.5$ and $\Theta=1$ as a function of the fictitious quantum statistics variable $\xi$ for various $N$. Red symbols: standard $\xi$-simulations from Ref.~\cite{Dornheim_JCP_xi_2023}; green (black): new $\xi$-simulations with the re-weighting estimator of Eq.~(\ref{eq:final}) for $\xi_\textnormal{ref}=1$ ($\xi_\textnormal{ref}=0.5$).
    }
\end{figure}

As the first example, we consider the UEG at $r_s=0.5$ and $\Theta=1$. These conditions correspond to strongly compressed states of matter as they are realized, e.g., at the NIF~\cite{Tilo_Nature_2023,Moses_NIF,Dornheim_NatComm_2025} and have also been considered for the original implementation of the $\xi$-extrapolation method into PIMC in Ref.~\cite{Dornheim_JCP_xi_2023}. In Fig.~\ref{fig:UEG_sign}, we show the average sign $S$ as a function of the fictitious quantum statistics variable $\xi$ on a semi-logarithmic scale. The red symbols have been obtained from individual, independent PIMC simulations for each given $\xi$ at different numbers of electrons $N$ and have been taken from Ref.~\cite{Dornheim_JCP_xi_2023}. They exhibit an exponential decrease with $\xi<0$, which is a consequence of the increasing number of pair permutations $N_\textnormal{pp}$, and which can be exploited, e.g., for the estimation of fermionic free energies~\cite{dornheim2025fermionicfreeenergiestextitab,svensson2025acceleratedfreeenergyestimation} using the recent $\eta$-ensemble framework~\cite{Dornheim_PRB_2025,Dornheim_PRR_2025}.

Let us next examine the green crosses and the black bars in Fig.~\ref{fig:UEG_sign}, which have been obtained from Eq.~(\ref{eq:final}) via a single PIMC simulation at reference values of $\xi_\textnormal{ref}=1$ and $\xi_\textnormal{ref}=0.5$, respectively. Both data sets are in excellent agreement with the reference data for every considered $N$, and over the full $\xi$-range, thereby fully validating the new re-weighting estimator, Eq.~(\ref{eq:final}). We note that our PIMC expectation values for $S$ vanish within the given error bars for $S\lesssim10^{-4}$.

\begin{figure}
    \centering
\includegraphics[width=1\linewidth]{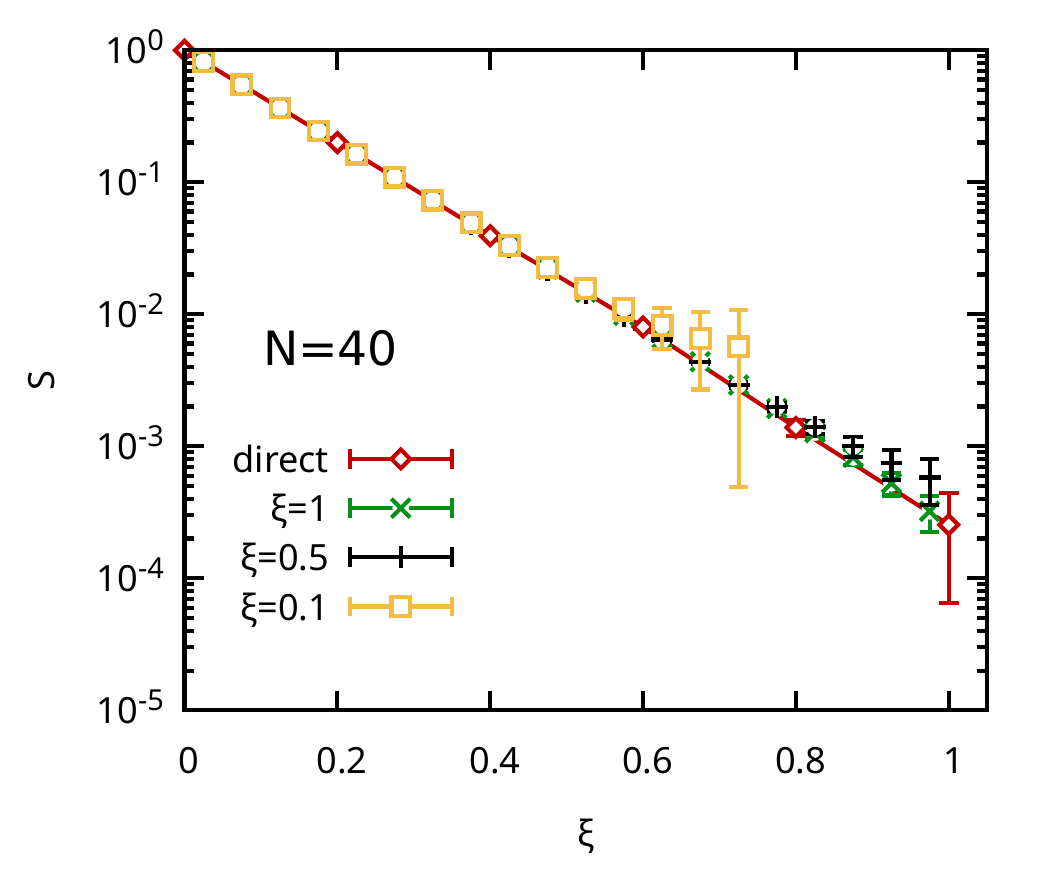}\\\vspace*{-1.1cm}\includegraphics[width=1\linewidth]{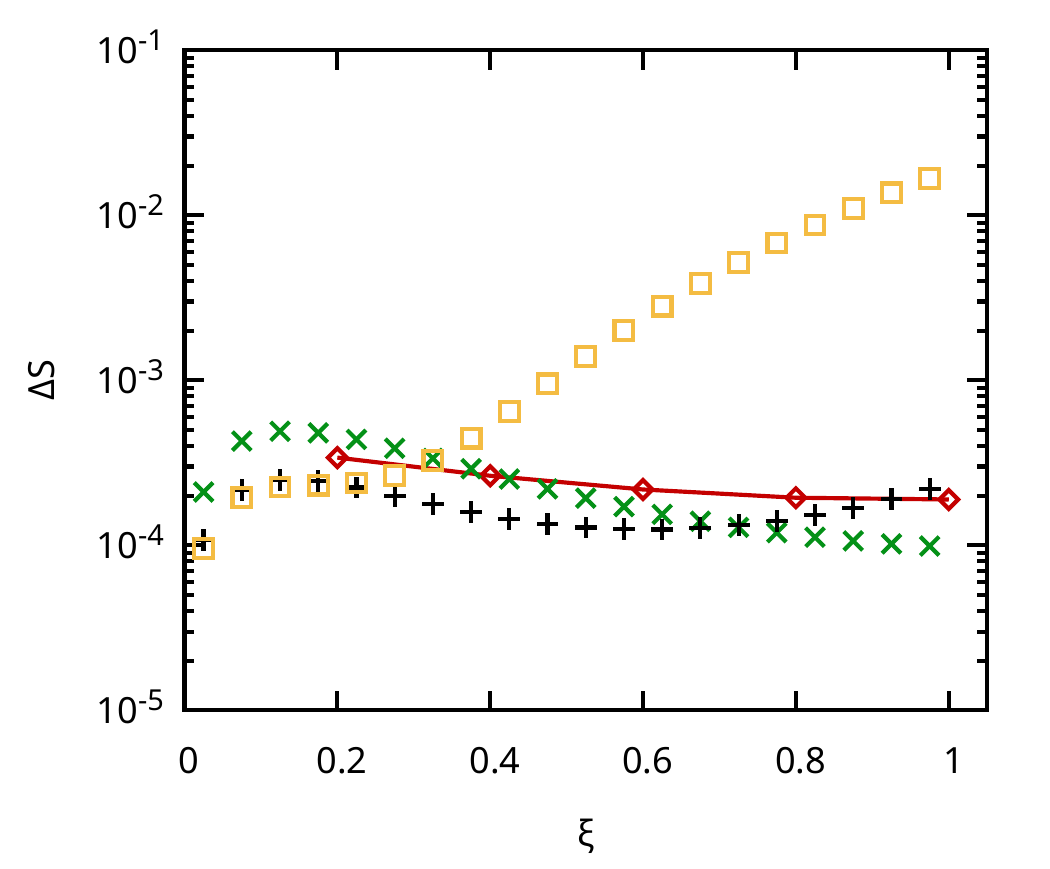}
    \caption{\label{fig:UEG_sign_N40} Top: average sign $S$ of the unpolarized UEG at $r_s=0.5$ and $\Theta=1$ as a function of the fictitious quantum statistics variable $\xi$ for $N=40$. Standard $\xi$-simulations from Ref.~\cite{Dornheim_JCP_xi_2023} (red symbols) together with new $\xi$-simulations with the re-weighting estimator of Eq.~(\ref{eq:final}) for $\xi_\textnormal{ref}=1$ (green symbols), $\xi_\textnormal{ref}=0.5$ (black symbols) and $\xi_\textnormal{ref}=0.1$ (yellow symbols). Bottom: corresponding statistical uncertainty.}
\end{figure}

Let us next consider the effect of different $\xi_\textnormal{ref}$, which is investigated in more detail in Fig.~\ref{fig:UEG_sign_N40}. The top panel shows the average sign for the case of $N=40$ with the green crosses, black bars and yellow squares corresponding to re-weighting results for $\xi_\textnormal{ref}=1,0.5,0.1$, respectively, and the red diamonds are direct PIMC reference results taken from Ref.~\cite{Dornheim_JCP_xi_2023}. First and foremost, we again find perfect agreement between all data sets over the entire depicted $\xi$-range, as it is expected.
At the same time, we find substantially larger error bars in the $\xi_\textnormal{ref}=0.1$ results for $|\xi|\gtrsim0.5$. This can be seen particularly well in the bottom panel of Fig.~\ref{fig:UEG_sign_N40}, where we show the absolute statistical uncertainty $\Delta S$ as a function of $\xi$. We note that the present $\xi_\textnormal{ref}$ calculations have been carried out using an equivalent amount of compute time, making them directly comparable with respect to the efficiency. The red diamonds are taken from Ref.~\cite{Dornheim_JCP_xi_2023} and have been obtained with a different set-up and varying computational resources; they are just included as a reference. Evidently, the choice of $\xi_\textnormal{ref}=1$ leads to the best results near the physical limit, $|\xi|\gtrsim0.8$, whereas $\xi_\textnormal{ref}=0.5,0.1$ are more efficient choices for $|\xi|\to0$. Overall, $\xi_\textnormal{ref}=0.5$ is a reasonable choice, since it provides good accuracy over the entire $\xi$-range.

\begin{figure}
    \centering
\includegraphics[width=1\linewidth]{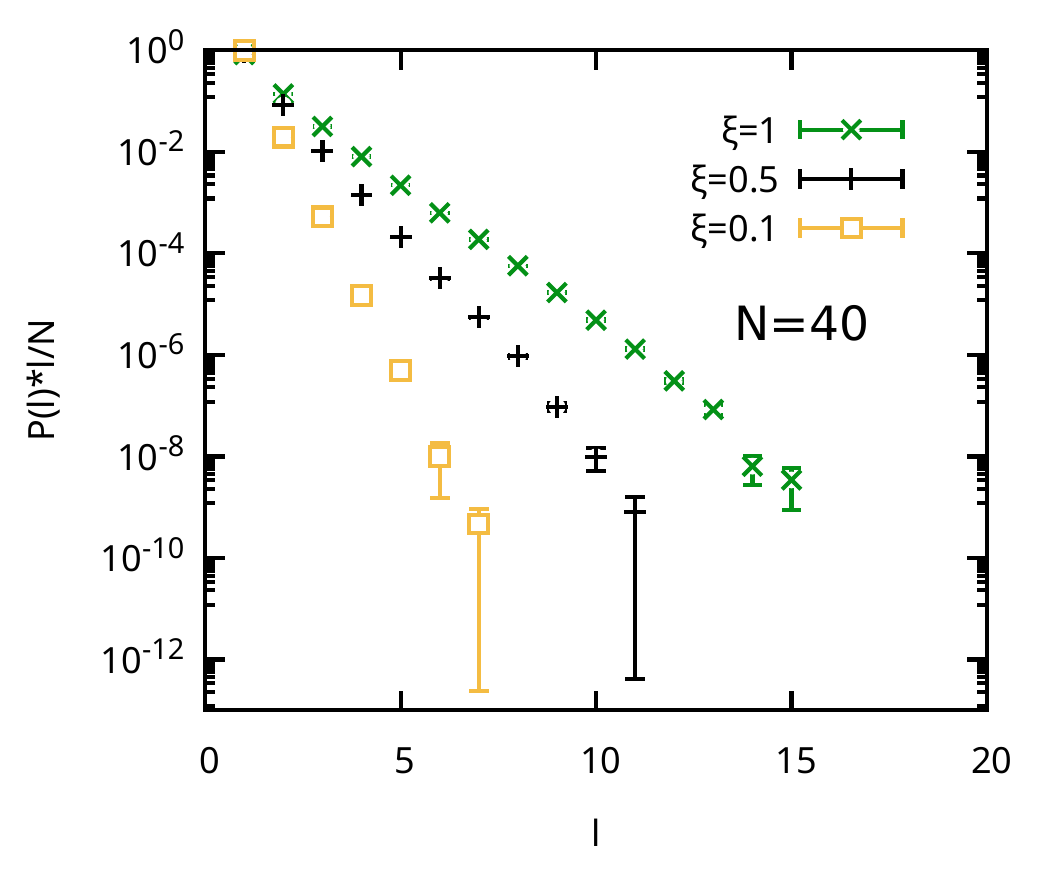}\caption{\label{fig:UEG_permutations_N40} Probability to find a particle in a permutation cycle of length $l$ for the unpolarized UEG at $r_s=0.5$ and $\Theta=1$ with $N=40$ and selected values of the fictitious quantum statistics variable $\xi$.
    }
\end{figure}

To get a more intuitive feeling for the re-weighting procedure and the dependence on $\xi_\textnormal{ref}$, we plot the probability to find a particle in a permutation cycle of length $l$~\cite{Dornheim_permutation_cycles} in Fig.~\ref{fig:UEG_permutations_N40} for all the three considered values of $\xi_\textnormal{ref}$. Note that we simulate an unpolarized electron gas here, making $l=20$ the largest possible permutation cycle length for a total particle number of $N=40$. In practice, we encounter cycles with a length of up to $l=15$ in our PIMC simulation, but their probability is at least eight orders of magnitude smaller compared to single particles without any exchange. Clearly, decreasing $\xi$ in Eq.~(\ref{eq:Z_generic}) penalizes configurations $\mathbf{X}$ with a larger number of pair exchanges $N_\textnormal{pp}$, see also the discussion in Ref.~\cite{Dornheim_JCP_xi_2023}. This gives one a direct picture of the different configuration spaces that are being sampled for different choices of $\xi_\textnormal{ref}$. For $\xi_\textnormal{ref}=0.1$, the PIMC simulation mostly encounters very small permutation cycles, which are being sampled with very high statistics; hence, the small error bars for $|\xi|\lesssim0.3$ in Fig.~\ref{fig:UEG_sign_N40}. On the other hand, those configurations with long permutation cycles are almost never encountered, which explains the rapidly increasing error for increasing values of $|\xi|$. In fact, such an extreme choice of $\xi_\textnormal{ref}$ might eventually lead to ergodicity problems, as the reference and re-weighted configuration spaces lose their overlap. In practice, this can be easily checked, e.g., by comparing (re-weighted) results for two distinctly different choices for $\xi_\textnormal{ref}$. Conversely, configurations $\mathbf{X}$ that consist predominantly of single particles are under-sampled for $\xi_\textnormal{ref}=1$, which explains the larger error bars of these calculations for $|\xi|\lesssim0.3$. Finally, the $\xi_\textnormal{ref}=0.5$ simulation covers both the single-particle and long-permutation cycle regimes reasonably well, leading to excellent results over the entire $\xi<0$ fermionic sector.


\begin{figure}
    \centering
\includegraphics[width=1\linewidth]{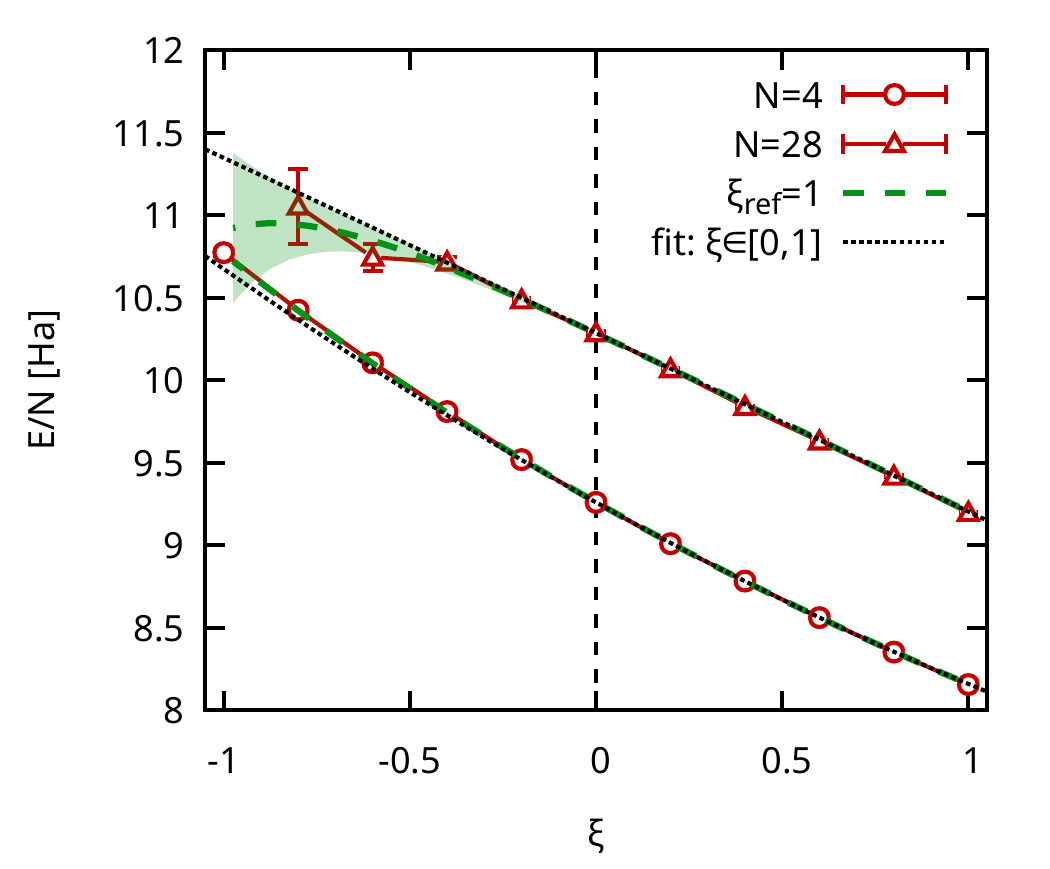}
\\\vspace*{-1.1cm}\includegraphics[width=1\linewidth]{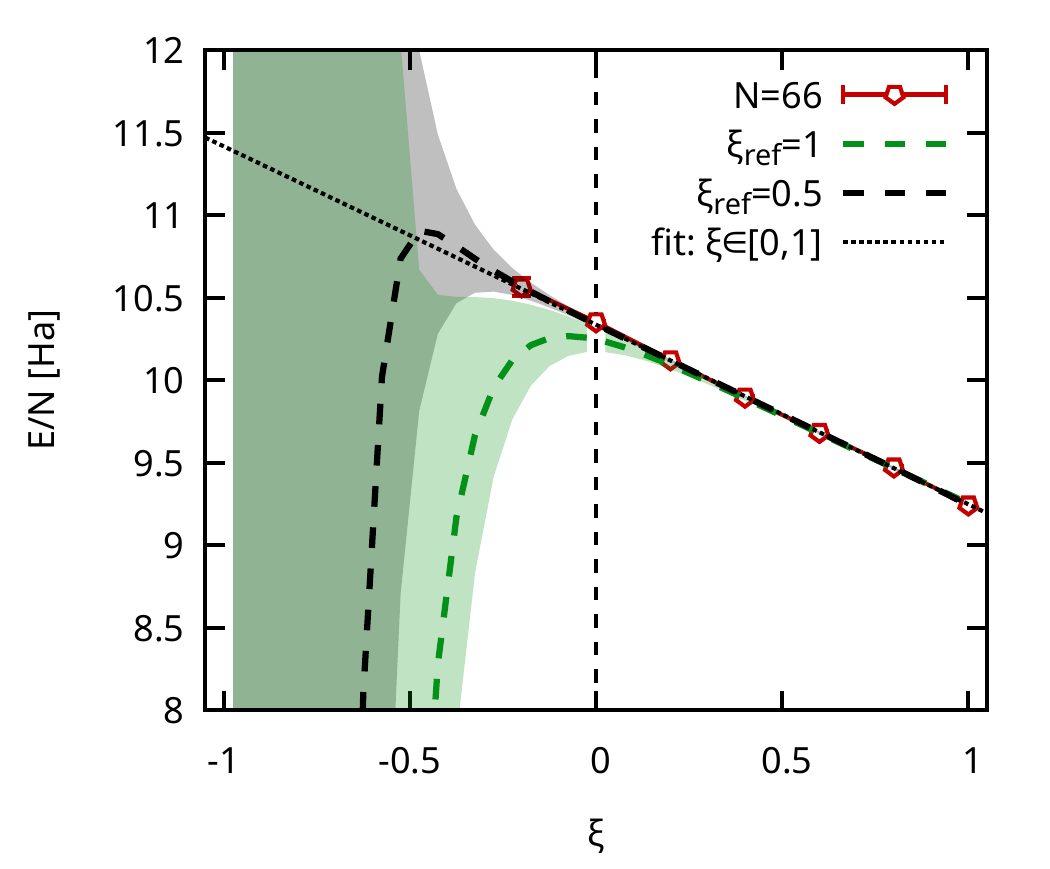}
    \caption{\label{fig:UEG_energy} Total energy per particle $E/N$ of the unpolarized UEG at $r_s=0.5$ and $\Theta=1$ as a function of the fictitious quantum statistics variable $\xi$ for various $N$. Red symbols: standard $\xi$-PIMC results taken from Ref.~\cite{Dornheim_JCP_xi_2023}; dashed green and dashed black: present work, 
    reweighted $\xi-$ PIMC results using our new estimator with $\xi_\textnormal{ref}=1$ and $\xi_\textnormal{ref}=0.5$, respectively, with the shaded areas indicating the statistical error bars. The dotted gray lines show quadratic fits via Eq.~(\ref{eq:extrapolation}) based on input from the sign-problem free domain of $\xi\geq0$.}
\end{figure}

Let us next come to the main application of the present work, which is given by the extrapolation of a physical observable $O(\xi)$ to the fermionic limit of $\xi=-1$. To this end, we follow Ref.~\cite{Dornheim_JCP_xi_2023} and consider the total energy per particle $E/N$, which is of great interest, e.g., for the construction of equation of state tables~\cite{Brown_PRB_2013,ksdt,groth_prl,review,Militzer_PRE_2021}.
In the top panel of Fig.~\ref{fig:UEG_energy}, we show the $\xi$-dependence of the total energy, with the red circles and triangles corresponding to standard $\xi-$PIMC reference values for $N=4$ and $N=28$, respectively. The dashed green lines show corresponding results using the present re-weighting estimator, Eq.~(\ref{eq:final}), with $\xi_\textnormal{ref}=1$ and the shaded green area depicts the associated statistical uncertainty. For $N=4$, the sign problem is very mild (see Fig.~\ref{fig:UEG_sign} above) and we find perfect agreement between the standard $\xi-$PIMC data set and the present re-weighted $\xi-$PIMC results despite the very small error bars. For $N=28$, the sign problem becomes substantially more severe towards the fermionic limit of $\xi=-1$, as reflected by the rapidly increasing error bars in the red triangles and the green curve. Nevertheless, we again find excellent agreement between the two data sets within the given error intervals. For completeness, we have also included quadratic fits following Eq.~(\ref{eq:extrapolation}) using input data from the sign-problem free domain of $\xi\geq0$, see the dotted black lines.

The bottom panel of Fig.~\ref{fig:UEG_energy} corresponds to $N=66$, for which the sign problem is very severe; indeed, the average sign cannot be resolved within the given Monte Carlo error bars for $\xi=-1$, cf.~Fig.~\ref{fig:UEG_sign} above. In this case, choosing $\xi_\textnormal{ref}=1$ leads to noticeably larger error bars around $\xi=0$, as this single-particle exchange-cycle regime is systematically under-sampled in the full PIMC calculation. 
Obviously, this is extremely detrimental for the $\xi$-extrapolation technique, which is based on accurately resolving the correct trend upon decreasing $\xi$. In stark contrast, the black curve showing the re-weighted $\xi-$PIMC results for $\xi_\textnormal{ref}=0.5$ is very accurate for $\xi\gtrsim-0.2$, with a comparable accuracy to the corresponding standard $\xi-$PIMC data point. Consequently, the subsequent $\xi$-extrapolation works very well, see the dotted gray curve.

\begin{figure}
    \centering
\includegraphics[width=1\linewidth]{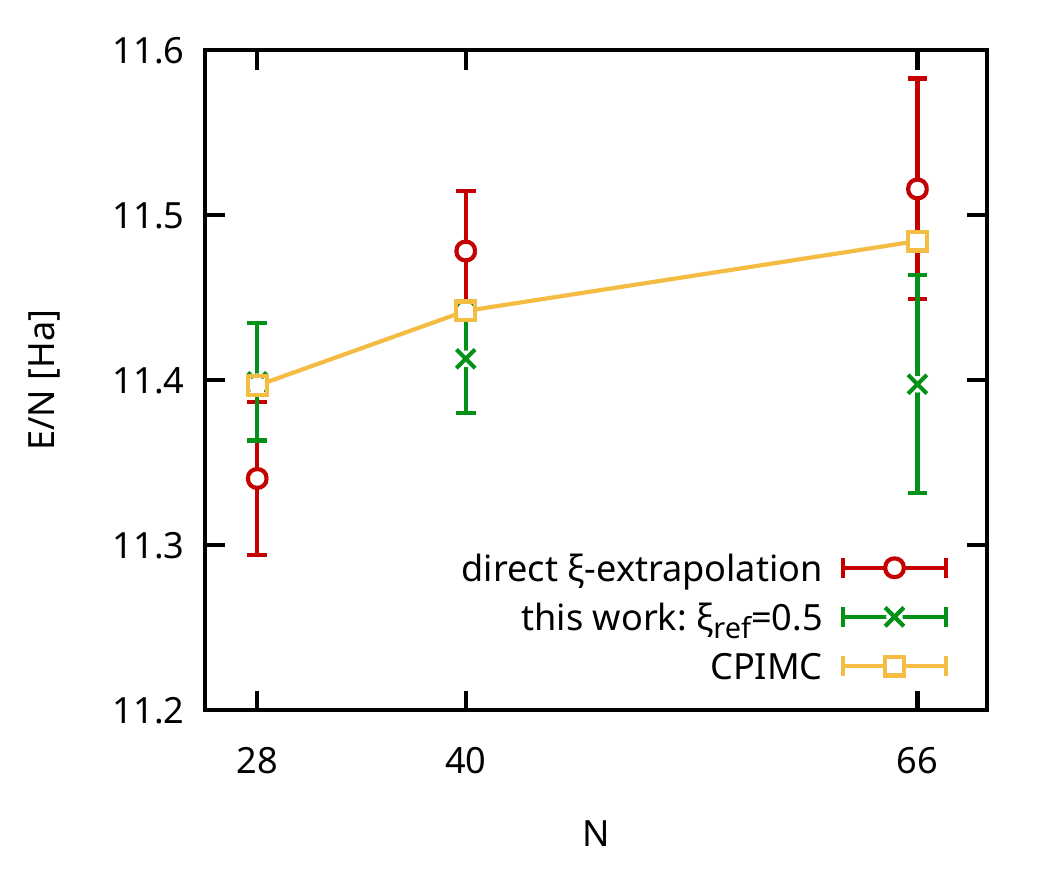}
    \caption{\label{fig:UEG_compare_energy} Total energy per particle $E/N$ of the unpolarized UEG at $r_s=0.5$ and $\Theta=1$ as a function of the system size in the fermionic limit $\xi=-1$. Yellow squares: exact CPIMC reference data; red circles: standard $\xi$-extrapolation, taken from Ref.~\cite{Dornheim_JCP_xi_2023}; green crosses: reweighted $\xi$-extrapolation with our new estimator [Eq.~(\ref{eq:final})] for $\xi_\textnormal{ref}=0.5$.
    }
\end{figure}

Let us conclude this investigation of the UEG by considering the extrapolated results at the fermionic limit $\xi=-1$ for the total energy per particle of the unpolarized UEG at $r_s=0.5$ and $\Theta=1$ for different particle numbers, as shown in Fig.~\ref{fig:UEG_compare_energy}.
The yellow squares show exact configuration PIMC (CPIMC) reference results with an error bar smaller than the symbol size. It is reminded that CPIMC is accurate for the present high density case, but breaks down with increasing coupling strength~\cite{Dornheim_POP_2017}. Moreover, it is pointed out that CPIMC simulations of realistic two-component systems that include both electrons and nuclei have not been demonstrated so far. The red circles have been obtained based on $N_\xi=6$ independent standard $\xi-$PIMC simulations for different values of $\xi$ and are taken from Ref.~\cite{Dornheim_JCP_xi_2023}. Finally, the green crosses have been obtained from a single reweighted $\xi-$PIMC simulation with our new re-weighting estimator [Eq.~(\ref{eq:final})] for $\xi_\textnormal{ref}=0.5$. They are in excellent agreement with the other two data sets within the given error bars, with the same order of magnitude in the statistical uncertainty compared to the standard $\xi$-PIMC simulation results. It is worth noting that corresponding direct PIMC calculations, that are subject to the full sign problem, result in an order of magnitude larger error bar for $N=28$ and do not converge at all for $N=40$ and $N=66$ given the same computational resources as the present approach.

\subsection{Strongly compressed beryllium\label{sec:beryllium}}

\begin{figure}
    \centering
\includegraphics[width=1\linewidth]{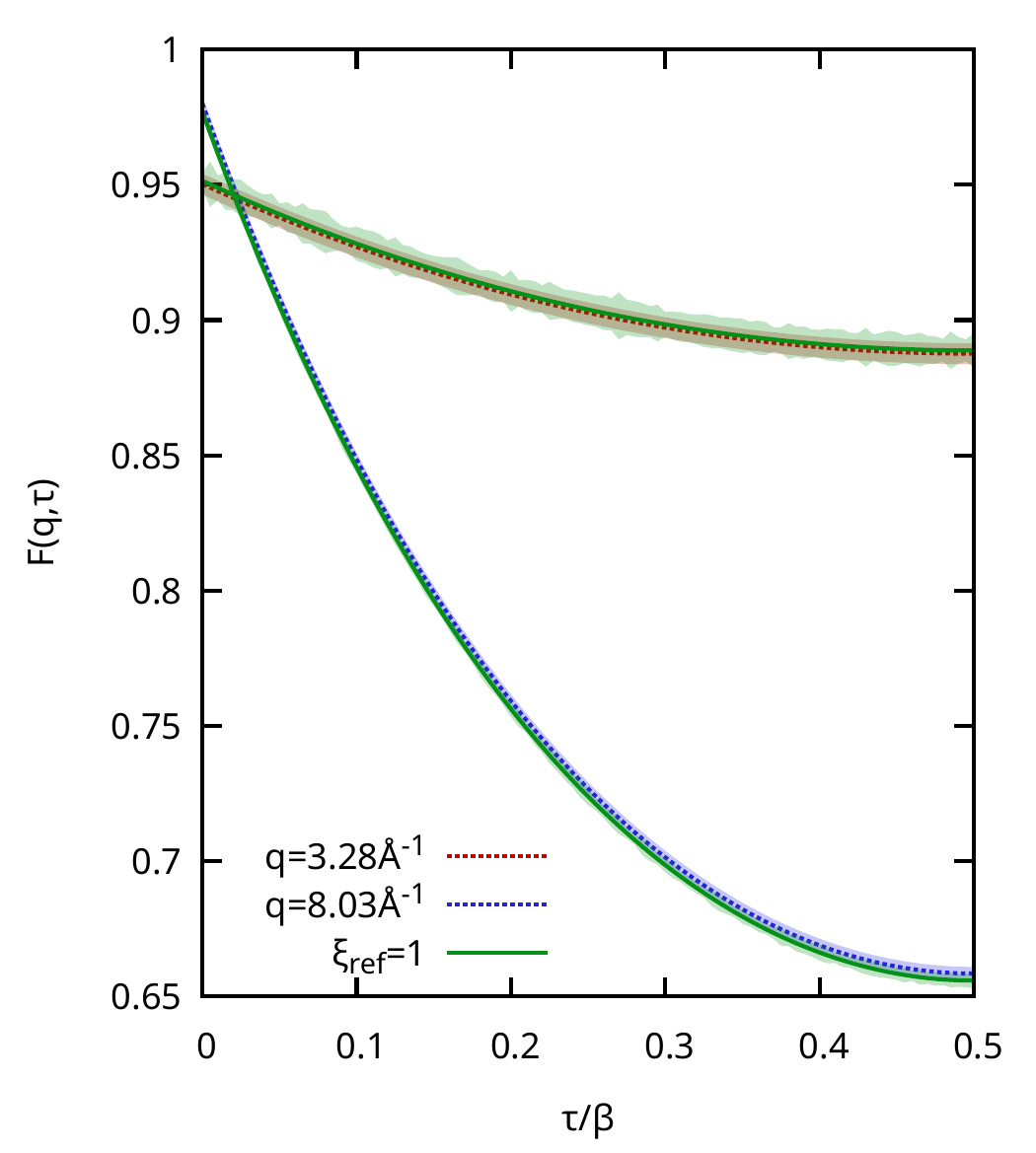}
    \caption{\label{fig:Be_ITCF} Imaginary-time correlation function (ITCF) $F(\mathbf{q},\tau)$ of compressed beryllium at $\rho=7.5\,$g/cc and $T=155.5\,$eV. Results at two selected wavenumbers for $N_\textnormal{Be}=10$ Be atoms. The red and blue curves correspond to direct sign-afflicted PIMC simulations and the green curves to $\xi-$PIMC simulations with the present re-weighting estimator [Eq.~(\ref{eq:final})] using $\xi_\textnormal{ref}=1$.
    }
\end{figure}

As the second and final example, we consider strongly compressed beryllium at a mass density of $\rho=7.5\,$g/cc ($r_s=0.93$) and a temperature of $T=155.5\,$eV ($\Theta=2.68$). Similar conditions have recently been realized at the NIF, in a spherical laser compression experiment by D\"oppner \emph{et al.}~\cite{Tilo_Nature_2023}.  Subsequently, Dornheim and co-workers have utilized the $\xi$-extrapolation technique to re-interpret these data sets~\cite{Dornheim_NatComm_2025,Dornheim_POP_2025,schwalbe2025staticlineardensityresponse}, which constitutes, to the best knowledge of the authors, the first direct comparison between a PIMC simulation and an x-ray scattering experiment. Specifically, the intensity that is measured in an x-ray Thomson scattering (XRTS) experiment is, to a very good approximation, given by the convolution of the electronic dynamic structure factor $S(\mathbf{q},\omega)$ with the combined source-and-instrument function $R(\omega)$~\cite{Dornheim_T_follow_up,Gawne_JAP_2025}
\begin{eqnarray}\label{eq:convolution}
    I(\mathbf{q},\omega) = S(\mathbf{q},\omega) \circledast R(\omega)\ .
\end{eqnarray}
The deconvolution of Eq.~(\ref{eq:convolution}) to extract $S(\mathbf{q},\omega)$ is notoriously unstable and, therefore, generally not feasible. Recently, Dornheim \emph{et al.}~\cite{Dornheim_T_2022,Dornheim_T_follow_up,Dornheim_SciRep_2024,Dornheim_POP_2025,Dornheim_MRE_2023,Dornheim_PTR_2023,schwalbe2025staticlineardensityresponse} have suggested to consider the imaginary-time density--density correlation function (ITCF) $F(\mathbf{q},\tau)=\braket{\hat{n}(\mathbf{q},0)\hat{n}(-\mathbf{q},\tau)}$, which is connected to $S(\mathbf{q},\omega)$ via a two-sided Laplace transform $\mathcal{L}\left[\dots\right]$,
\begin{eqnarray}\label{eq:deconvolution}
F(\mathbf{q},\tau) &=&     \mathcal{L}\left[S(\mathbf{q},\omega)\right] = \int_{-\infty}^\infty \textnormal{d}\omega\ S(\mathbf{q},\omega)\ e^{-\beta\omega} \\
&=& \frac{\mathcal{L}\left[S(\mathbf{q},\omega)\circledast R(\omega)\right]}{\mathcal{L}\left[R(\omega)\right]}\ .
\end{eqnarray}
For the second line, we have made use of the usual convolution theorem, which, together with the high robustness of $\mathcal{L}\left[\dots\right]$ with respect to experimental noise and accurate knowledge of $R(\omega)$, ensures that we can directly obtain the physical ITCF from an XRTS measurement. First and foremost, this procedure allows for a straightforward model-free extraction of a number of system parameters, such as the temperature~\cite{Dornheim_T_2022,Dornheim_T_follow_up,Schoerner_PRE_2023,shi2025firstprinciplesanalysiswarmdense}, absolute intensity~\cite{Dornheim_SciRep_2024,Dornheim_NatComm_2025} and degree of non-equilibrium~\cite{vorberger2023revealing,Bellenbaum_APL_2025}, which are of key importance in their own right. Second, Eq.~(\ref{eq:deconvolution}) potentially facilitates the direct comparison between XRTS and PIMC in the imaginary-time domain, see Ref.~\cite{Dornheim_NatComm_2025} for a first demonstration of this idea. Finally, the ITCF contains, by definition, the same information as $S(\mathbf{q},\omega)$, only in a different representation~\cite{Dornheim_MRE_2023,Dornheim_PTR_2023}; this insight has recently been exploited by Bellenbaum \emph{et al.}~\cite{Bellenbaum_PRR_2025}, who have used PIMC results for the ITCF of warm dense hydrogen to extract ionization states and the ionization potential depression. To summarize, the ITCF constitutes a key property uniting structural and spectral information, making it a well suited observable for the present study.

In Fig.~\ref{fig:Be_ITCF}, we show the ITCF of strongly compressed beryllium along the $\tau$-axis for two representative wavenumbers corresponding to XRTS experiments in a forward ($q=3.28$\AA$^{-1}$) and backward ($q=8.03$\AA$^{-1}$) scattering geometry. We note that the detailed balance relation of the frequency domain, that is valid in thermal equilibrium, manifests as a translation symmetry $F(\mathbf{q},\tau)=F(\mathbf{q},\beta-\tau)$ in the imaginary time domain, so we restrict ourselves to the non-redundant interval of $\tau\in[0,\beta/2]$. The red and the blue curves have been obtained from direct sign-afflicted PIMC simulations at $\xi=-1$, for which we find an average sign of $S=0.097$; consequently, such exact simulations are computationally involved, but still feasible on modern high-performance computing systems. Evidently, we find an increased $\tau$-decay with increasing $q$, which is a direct manifestation of the more pronounced impact of quantum delocalization on smaller length scales $\lambda=2\pi/q$, see Refs.~\cite{Dornheim_PTR_2023,Dornheim_MRE_2023} for a more detailed discussion of the physical interpretation of $F(\mathbf{q},\tau)$. The green curves show corresponding results combining the $\xi$-extrapolation technique [Eq.~(\ref{eq:extrapolation})] with our new re-weighting estimator [Eq.~(\ref{eq:final})]. We find excellent agreement between the present work and the exact sign-afflicted reference data within the given error bars over the entire $\tau$-range, as expected.

\begin{figure}
    \centering
\includegraphics[width=1\linewidth]{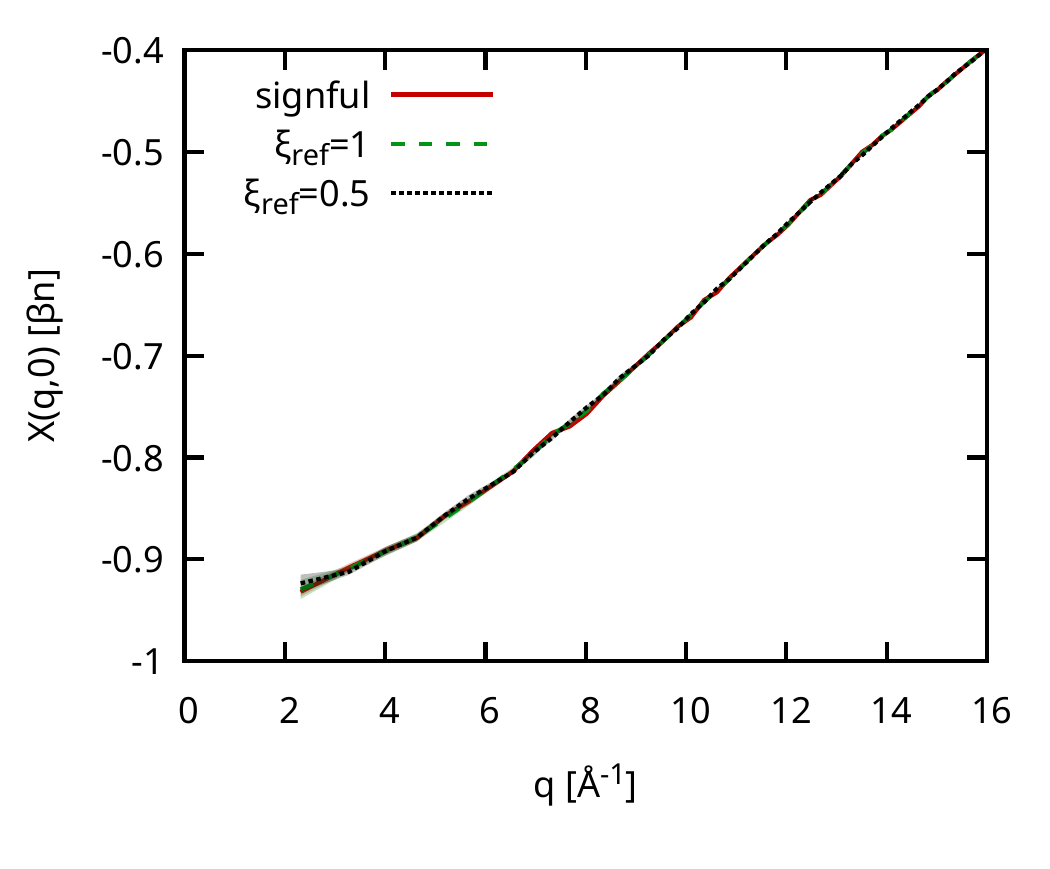}
    \caption{\label{fig:Be_IntFive} Static linear density response $\chi(\mathbf{q},0)$ [Eq.~(\ref{eq:static_chi})] of compressed beryllium at $\rho=7.5\,$g/cc and $T=155.5\,$eV. Solid red: exact sign-afflicted PIMC reference data; dashed green (dotted black): our new $\xi-$PIMC data with the present re-weighting estimator [Eq.~(\ref{eq:final})] using $\xi_\textnormal{ref}=1$ ($\xi_\textnormal{ref}=0.5$).}
\end{figure}

As the final property, let us consider the static (i.e., $\omega\to0$) limit of the dynamic linear density response function $\chi(\mathbf{q},\omega)$, which is connected to the ITCF via the imaginary-time version of the fluctuation--dissipation theorem:
\begin{eqnarray}\label{eq:static_chi}
    \chi(\mathbf{q},0) = -n \underbrace{\int_0^\beta\textnormal{d}\tau\ F(\mathbf{q},\tau)}_{L(\mathbf{q})}\ .
\end{eqnarray}
Eq.~(\ref{eq:static_chi}) has been used for extensive PIMC studies of the static local field factor both of the UEG~\cite{dornheim_ML,dynamic_folgepaper,Tolias_JCP_2021,dornheim_electron_liquid,Dornheim_PRB_ESA_2021,Dornheim_PRL_2020_ESA} and, very recently, of warm dense hydrogen~\cite{Dornheim_MRE_2024}.
In addition, Schwalbe \emph{et al.}~\cite{schwalbe2025staticlineardensityresponse} have suggested to use the area under the ITCF $L(\mathbf{q})$ as a diagnostic for the mass density from XRTS measurements. In Fig.~\ref{fig:Be_IntFive}, we show corresponding PIMC results for the wavenumber dependence of $\chi(\mathbf{q},0)$, with the red curve again depicting the exact sign-afflicted PIMC reference data set. The dashed green and dotted black curves have been obtained based on $\xi-$PIMC simulations with our new re-weighting estimator, Eq.~(\ref{eq:final}), for $\xi_\textnormal{ref}=1$ and $\xi_\textnormal{ref}=0.5$, respectively. They are in excellent agreement with the reference data over the entire range of wavenumbers, thereby further highlighting the versatility and general applicability of our new set-up.

\section{Summary and Outlook\label{sec:outlook}}


We have presented a new re-weighting estimator for PIMC simulations of fictitious identical particles that allows to obtain the full $\xi$-dependence from a single simulation of an, in principle, arbitrary reference value $\xi_\textnormal{ref}$. This removes the need for $N_\xi=10-20$ independent PIMC simulations, which were required in previous applications of the $\xi$-extrapolation technique~\cite{Xiong_JCP_2022,Dornheim_JCP_xi_2023,Dornheim_JPCL_2024,Dornheim_JCP_2024} to obtain the fermionic limit of $\xi=-1$. To demonstrate the general applicability of the new estimator, we have applied it to two representative examples: the energy per particle of the warm dense electron gas at $r_s=0.5$ and $\Theta=1$ as well as the imaginary time correlation function and static linear density response of strongly compressed beryllium with $\rho=7.5\,$g/cc and $T=155.5\,$eV. In both cases, we were able to reproduce the earlier PIMC results~\cite{Dornheim_JCP_xi_2023,Dornheim_NatComm_2025} with a substantially reduced computational effort.

Due to its general nature, our idea can be used in a gamut of future PIMC simulations of partially degenerate quantum many-fermion systems, including electrons in quantum dots, ultracold atoms as well as a variety of warm dense matter systems. In addition, we note that our idea can easily be adapted to path integral molecular dynamics simulations~\cite{Xiong_JCP_2022,xiong2024gpuaccelerationabinitio,Hirshberg_JCP_2020,Hirshberg_PNAS_2019}, for which the $\xi$-extrapolation technique was originally developed.

\begin{acknowledgements}

\noindent This work was partially supported by the Center for Advanced Systems Understanding (CASUS), financed by Germany’s Federal Ministry of Education and Research and the Saxon state government out of the State budget approved by the Saxon State Parliament. This work has received funding from the European Research Council (ERC) under the European Union’s Horizon 2022 research and innovation programme (Grant agreement No. 101076233, "PREXTREME"). 
Views and opinions expressed are however those of the authors only and do not necessarily reflect those of the European Union or the European Research Council Executive Agency. Neither the European Union nor the granting authority can be held responsible for them.
Computations were performed on a Bull Cluster at the Center for Information Services and High-Performance Computing (ZIH) at Technische Universit\"at Dresden and at the Norddeutscher Verbund f\"ur Hoch- und H\"ochstleistungsrechnen (HLRN) under grant mvp00024.
\end{acknowledgements}

\bibliography{bibliography}

\begin{thebibliography}{120}
\bibitem [{rep()}]{repo}%
  \BibitemOpen
  \href@noop {} {}\bibinfo {note} {A link to a repository containing all PIMC results will be made available upon publication.}\BibitemShut {Stop}%
\end{thebibliography}
\end{document}